\documentclass[journal=nanoletters,manuscript=letter]{achemso}
\usepackage{textcomp}
\usepackage{multirow}
\usepackage{threeparttable}
\usepackage{dcolumn}
\usepackage{bm}
\usepackage{mathtools,amssymb}
\usepackage{upgreek}
\usepackage{siunitx}
\usepackage{color}
\usepackage{float}
\usepackage{textcomp}
\usepackage[utf8]{inputenc}
\usepackage{mathptmx}
\usepackage{etoolbox}
\usepackage[symbol]{footmisc}
\usepackage{chemformula} 
\usepackage[T1]{fontenc} 
\usepackage{textgreek} 
\usepackage{amsmath}
\usepackage{textcomp}

\makeatletter

\author{Prasenjit Ghosh}
\affiliation[Indian Institute of Technology, Kharagpur]
{Materials Science Centre, Indian Institute of Technology, Kharagpur, West Bengal 721302, India}

\author{Sourav Paul}
\affiliation[Indian Institute of Technology, Kharagpur]
{Materials Science Centre, Indian Institute of Technology, Kharagpur, West Bengal 721302, India}
\altaffiliation{Present address: Department of Condensed Matter and Materials Physics, S. N. Bose National Centre for Basic Sciences, Sector III, Block JD, Salt Lake, Kolkata 700106, India}

\author{Subhobrata Banerjee}
\affiliation[Indian Institute of Technology, Kharagpur]
{School of Nanoscience and Technology, Indian Institute of Technology, Kharagpur, West Bengal 721302, India}

\author{Abhijith M B}
\affiliation[Indian Institute of Technology, Kharagpur]
{Materials Science Centre, Indian Institute of Technology, Kharagpur, West Bengal 721302, India}

\author{Vineet Pandey}
\affiliation[Indian Institute of Technology, Kharagpur]
{Materials Science Centre, Indian Institute of Technology, Kharagpur, West Bengal 721302, India}
\altaffiliation{Present address: Centre for Nano Science and Engineering, Indian Institute of Science, Bangalore 560012, India}

\author{Premananda Chatterjee}
\affiliation[S. N. Bose National Centre for Basic Sciences, Kolkata]
{Department of Condensed Matter and Materials Physics, S. N. Bose National Centre for Basic Sciences, Sector III, Block JD, Salt Lake, Kolkata 700106, India}

\author{Kenji Watanabe}
\affiliation[National Institute for Materials Science]
{Research Center for Electronic and Optical Materials, National Institute for Materials Science, 1-1 Namiki, Tsukuba 305-0044, Japan}

\author{Takashi Taniguchi}
\affiliation[National Institute for Materials Science]
{Research Center for Materials Nanoarchitectonics, National Institute for Materials Science, 1-1 Namiki, Tsukuba 305-0044, Japan}

\author{Ajit Roy}
\affiliation[Air Force Research Laboratory]
{Air Force Research Laboratory, Wright-Patterson Air Force Base, OH 45433, United States}

\author{Nicholas R Glavin}
\affiliation[Air Force Research Laboratory]
{Air Force Research Laboratory, Wright-Patterson Air Force Base, OH 45433, United States}

\author{Atindra Nath Pal}
\affiliation[S. N. Bose National Centre for Basic Sciences, Kolkata]
{Department of Condensed Matter and Materials Physics, S. N. Bose National Centre for Basic Sciences, Sector III, Block JD, Salt Lake, Kolkata 700106, India}

\author{Krishna Prasad Maity}
\email{krishnaprasad.m@srmap.edu.in}
\affiliation[SRM University AP, Andhra Pradesh]
{Department of Physics, SRM University AP, Andhra Pradesh 522240, India}

\author{Vidya Kochat}
\email{vidya@matsc.iitkgp.ac.in}
\affiliation[Indian Institute of Technology, Kharagpur]
{Materials Science Centre, Indian Institute of Technology, Kharagpur, West Bengal 721302, India}

\title{Electrostatic Boundary Engineering for Room-Temperature Ferroelectric Retention in $\alpha$-In$_2$Se$_3$/Graphene Neuromorphic Devices}

\begin{document}

\maketitle

\begin{abstract}
2D ferroelectrics such as $\alpha$-In$_2$Se$_3$ offers a promising platform for neuromorphic computing due to switchable polarization states, that enable electrically programmable conductance states. However, the polarization retention capability of such states in ultrathin $\alpha$-In$_2$Se$_3$ layers depends on factors such as temperature and dielectric environment, which needs to be benchmarked. Here we investigate the temperature-dependent stability of polarization-mediated resistance states in $\alpha$-In$_2$Se$_3$/graphene heterostructures under distinct electrostatic boundary conditions. In the weakly screened configuration, the polarization-induced resistance states in graphene exhibit rapid relaxation at room temperature but become remarkably stable upon cooling below 100~K.  In contrast, incorporating a graphite electrode modifies the electrostatic screening at the ferroelectric-metal interface and markedly enhances the polarization retention at room temperature. These enhanced retention characteristics preserve the programmed conductance states over longer timescales and improve the probability of MNIST handwritten digit recognition, highlighting the importance of electrostatic boundary engineering in determining neuromorphic performance. 

\end{abstract}


Two-dimensional (2D) van der Waals (vdW) ferroelectrics have emerged as a promising platform for ultrathin, low-power, and nonvolatile electronics.\cite{Wang2023,Song2026,Zhang2022,Wang2025,Liu2021,Scott2007,Khan2020} Unlike conventional ferroelectrics, where depolarization fields increasingly destabilize polarization at reduced thickness, vdW ferroelectrics can sustain spontaneous polarization down to few-layer and monolayer limits. \cite{Chang2016,Guo2025,Liu2016,Xue2018,Xue2018_2,Paul2026,Yasuda2021,Yasuda2024,Wang2022,Jiang2024,Yang2023,Zheng2020} Their dangling bond-free surfaces also enable atomically clean integration with other vdW materials. 2D $\alpha$-In$_2$Se$_3$ is particularly attractive due to its coupled in-plane and out-of-plane polarizations and have been explored for neuromorphic applications, including pulse-driven conductance modulation and synaptic plasticity. \cite{Ding2017,Cui2018,Xue2019,Zhou2017,Xu2020,Huang2022,Wang2020,Xue2018_2,Wan2019} Furthermore, non-volatile memory behaviour and polarization-dependent electronic transport have been demonstrated in graphene/$\alpha$-In$_2$Se$_3$ heterostructures, where the ferroelectric nature of $\alpha$-In$_2$Se$_3$ was electrically probed through the polarization-induced shift of the Fermi level in graphene. \cite{Wan2019}

However, achieving stable polarization in practical devices remains challenging due to the strong influence of the electrostatic environment on the  polarization retention, through charge transfer and screening at the ferroelectric–electrode interface. \cite{Li2018screening,Junquera2003} Recent works have highlighted the importance of interfacial charge screening in controlling the polarization response of 2D ferroelectric heterostructures. \cite{Huang2024,Nahid2024}.  In graphene/In$_2$Se$_3$/graphene heterostructures, the incompletely screened polarization charges was found to result in a large depolarization field. \cite{Ding2017,Wan2019,Nahid2024} Polarization switching or the observation of a hysteresis window alone does not establish a long-term stability of a programmed conductance state. The stability of the polarization state is critically influenced by electrostatic boundary conditions and domain relaxation kinetics. For neuromorphic computing, the devices must support gradual and controllable conductance updates during learning while maintaining sufficient programmed state stability during inference.\cite{Joo2023,Huang2025} Uncontrolled relaxation can lead to synaptic weight drift and compromise inference reproducibility, whereas tunable volatility may be advantageous for short-term memory and temporal information processing. A systematic investigation of retention dynamics and interfacial charges is therefore essential to establish stable programmable conductance states in artificial synaptic devices based on 2D vdW heterostructures.

Here, we investigate the polarization retention characteristics of conductance states in $\alpha$-In$_2$Se$_3$/ hBN/ graphene ferroelectric field-effect transistors (Fe-FETs) by probing the Fermi level shifts in graphene, an established sensitive approach to monitor the switching dynamics of underlying $\alpha$-In$_2$Se$_3$. We observed that the switching behavior is governed by the interplay between depolarization-field screening and thermal fluctuations. Unstable polarization states were observed at room temperature under weak dielectric screening, while reducing the temperature progressively suppressed polarization relaxation and led to stable states below 100~K despite the persistence of a depolarization field. In contrast, introducing a graphite electrode directly above $\alpha$-In$_2$Se$_3$ screens the depolarization field leading to robust polarization states at room temperature. These results establish electrostatic boundary engineering as a key strategy for controlling polarization retention in 2D ferroelectrics, providing a design route towards room temperature-stable and low-power devices for neuromorphic computing.

\begin{figure*}
\includegraphics[width=1\textwidth]{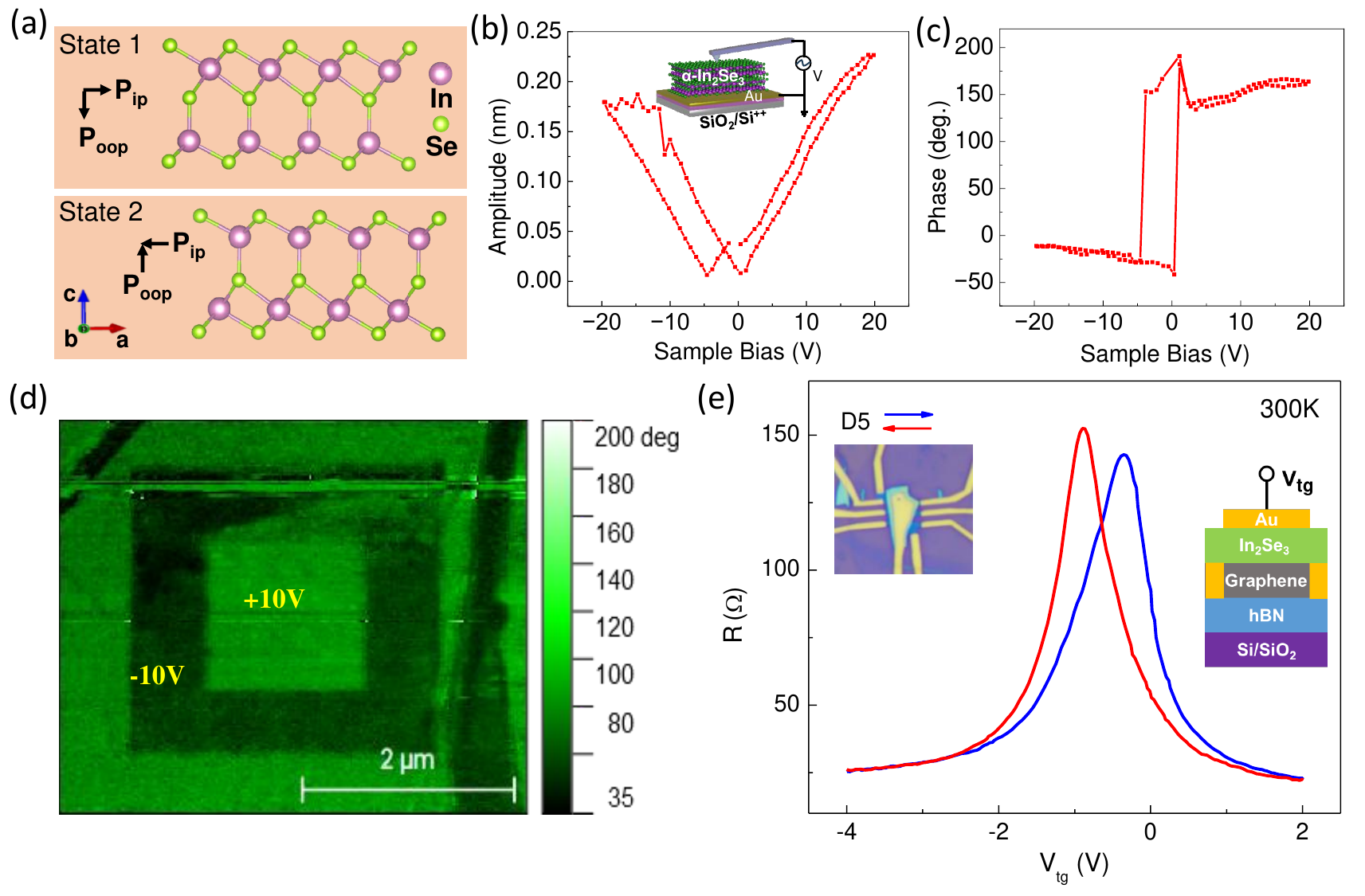}
\caption{(a) Atomic arrangement in $\alpha$-In$_2$Se$_3$ for the two polarization states 1 and 2. (b, c) PFM amplitude and phase hysteresis loops, indicating the ferroelectric behavior of $\alpha$-In$_2$Se$_3$. (d) PFM phase area map of $\alpha$-In$_2$Se$_3$ demonstrating oppositely polarized ferroelectric domains written under applied DC bias of +10~V and -10~V. (e) R-V$_{tg}$ characteristics of graphene channel in Graphene FeFET with $\alpha$-In$_2$Se$_3$ as the gate dielectric.} \label{Fig1}
\end{figure*}

A single unit cell thick $\alpha$-In$_2$Se$_3$ consists of a quintuple layer of alternating In and Se atoms with non-centrosymmetric R3m space-group symmetry. \cite{Ding2017,Zhou2017,Raman_In2Se3_MRB,Dutta2021,Liu2019} This symmetry breaking results in a spontaneous polarization both in-plane (P$_{ip}$) and in the out-of-plane (P$_{oop}$) directions, as shown in Fig.~\ref {Fig1}a. Ultra-thin $\alpha$-In$_2$Se$_3$ films were micromechanically exfoliated from bulk crystals and transferred onto a gold-deposited SiO$_2$/Si$^{++}$ wafer to investigate the ferroelectric nature through piezo force microscopy (PFM) studies. \cite{Gao2021,Li2020} Under an applied DC sample bias across $\alpha$-In$_2$Se$_3$, the PFM amplitude displays a butterfly loop with minima at coercive voltages of $+0.5$~V and $-4.57$~V as shown in Fig.~\ref{Fig1}b, which arises from the polarization switching between states 1 and 2 (shown in Fig.~\ref{Fig1}a).\cite{Sahota2004,Zhou2017} The large piezo-response at high bias voltages, together with a 180$^\circ$ out-of-plane phase contrast in the PFM phase loop shown in Fig.~\ref{Fig1}c, confirms the switchable polarization of $\alpha$-In$_2$Se$_3$ domains along the $c$-axis. The PFM phase map of oppositely poled domains generated by sequential $\pm10$~V tip biases is shown in Fig.~\ref{Fig1}d. Initially, a $2.5\times2.5\ \mu m^2$ square region was biased to state 1 by $-10$~V and a nested $1\times1\ \mu m^2$ region was switched to state 2 by $+10$~V bias. The sharp phase contrast in the subsequent zero-bias scans demonstrates retention of the polarization even after 2000~s. Following this, a graphene/In$_2$Se$_3$ Fe-FET with an Au top gate was fabricated, as illustrated in the schematic in Fig.~\ref{Fig1}e. The graphene channel transfer characteristics shown in Fig.~\ref{Fig1}e, exhibited a pronounced Dirac-point shift between the forward and reverse top gate voltage ($V_{tg}$) sweeps, indicative of polarization switching in $\alpha$-In$_2$Se$_3$ layer.\cite{Wan2019,Ghani2024,Paul2026}

\begin{figure*}
\includegraphics[width=1\textwidth]{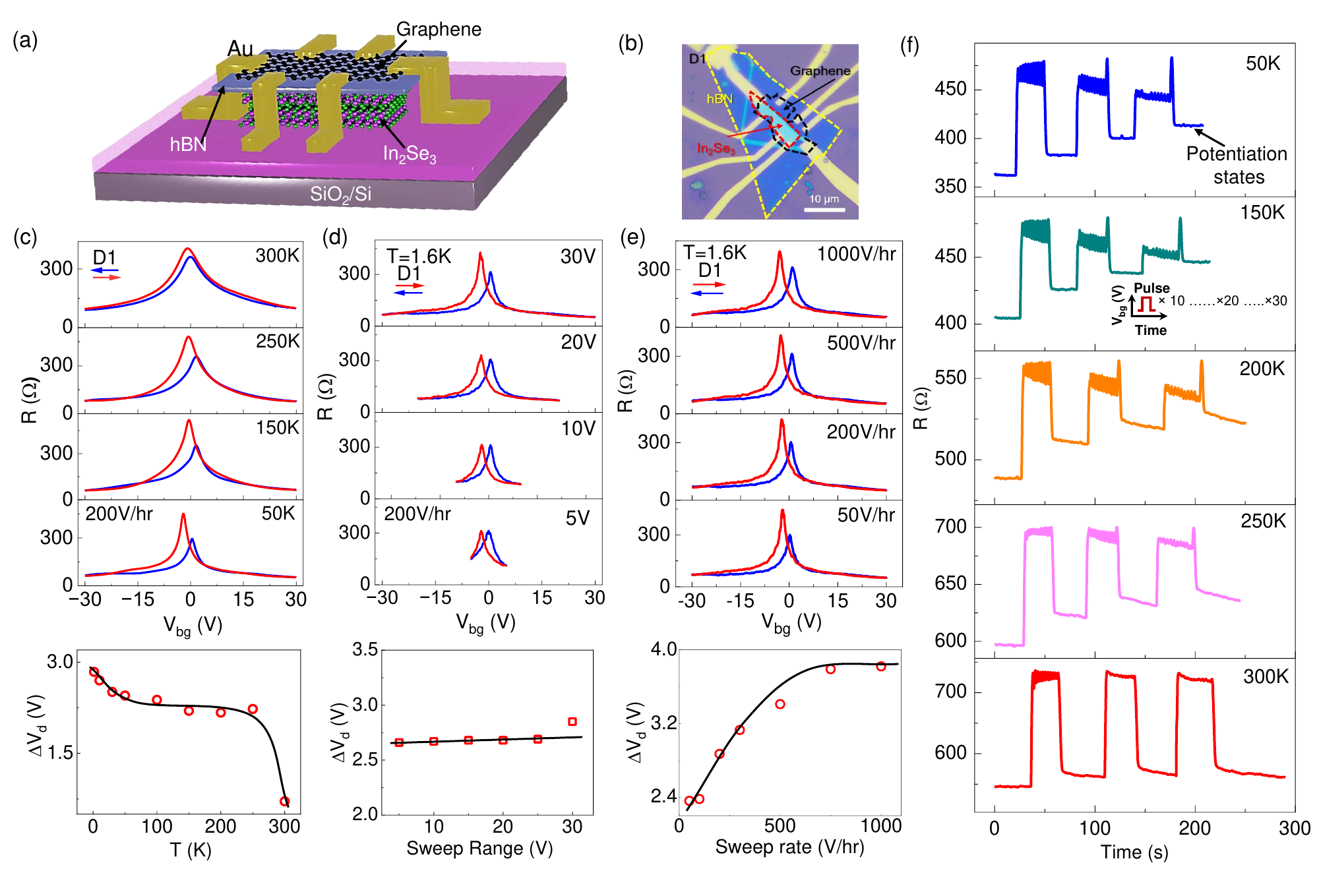}
\caption{(a) Schematic of $\alpha$-In$_2$Se$_3$/hBN/Graphene heterostructure on SiO$_2$/Si substrate for Graphene Fe-FET device. (b) Optical micrograph of the Graphene Fe-FET device D1. (c) Top panel: R-V$_{bg}$ curve for device D1 at different temperatures. Bottom panel: $\Delta V_d$ as a function of temperature showing an increase in the value of $\Delta V_d$ at low temperatures. (d) Top panel: R-V$_{bg}$ curve for device D1 at 1.6~K for different gate-voltage ranges. Bottom panel: $\Delta V_d$ as a function of V$_{bg}$ sweep range.  (e) Top panel: R-V$_{bg}$ curve for device D1 at 1.6~K at different sweep rates. Bottom panel: $\Delta V_d$ as a function of V$_{bg}$ sweep rate. (f) The resistance states for device D1 under  controlled voltage pulses at 50~K, 150~K, 200~K, 250~K, and 300~K, respectively. The inset shows the pulse sequence for potentiation measurements. } \label{Fig2}
\end{figure*}

To evaluate their potential for energy-efficient neuromorphic computing, we next investigated the stability, retention lifetime, and operating voltages of Graphene/hBN/$\alpha$-In$_2$Se$_3$ Fe-FETs. A typical device architecture is schematically illustrated in Fig.~\ref{Fig2}a, where graphene serves as an electrical transducer, offering remarkable sensitivity to examine the polarization switching characteristics of $\alpha$-In$_2$Se$_3$. A thin spacer layer of hBN is introduced to reduce the charge inhomogeneity in graphene channel caused by the direct influence of interface disorder or structural defects in In$_2$Se$_3$.\cite{Dean2010,Park2015,Ghani2024} Ferroelectric switching in $\alpha$-In$_2$Se$_3$ was studied using five Hall-bar Graphene Fe-FETs with (D1, D3, D4) and without (D2, D5) hBN spacers. Fig.~\ref{Fig2}b shows the optical micrograph of device D1 whose transfer characteristics obtained on varying the voltage through Si$^{++}$ back gate ($V_{bg}$) in both forward and reverse sweeps is indicated in Fig.~\ref{Fig2}c (top panel) for different temperatures. The evolution of four-probe resistance of graphene in device D1 at 1.6~K  for different $V_{bg}$ scan ranges swept at a constant rate of 200~V/hr is demonstrated in Fig.~\ref{Fig2}d (top panel). The Dirac-point shift between the forward and reverse gate voltage sweeps, $\Delta V_d$, indicates the additional charge carrier density ($\Delta \sigma_p$) induced in graphene from the polarization of $\alpha$-In$_2$Se$_3$ under gate electric field, and is given by \cite{Wan2019,Paul2026,Ghani2024}
\begin{equation}
    \Delta \sigma_p = C_{tot} \times {\Delta V_d }
\end{equation}\label{eqn1}
where $C_{tot} = \frac{1}{\frac{1}{C_{SiO_2}}+\frac{1}{C_{In_2Se_3}}+\frac{1}{C_{hBN}}}$, represents the total capacitance across the graphene channel and gate electrode. Varying the $V_{bg}$ sweep limits under a constant ramp rate of 200~V/hr did not modify $\Delta V_d$ (bottom panel of Fig.~\ref{Fig2}d), suggesting saturation of switchable dipoles near the h-BN / In$_2$Se$_3$ interface under very low threshold voltages. On the other hand, systematic variation of $V_{bg}$ sweep rate within a fixed voltage window of $\pm30$~V led to a progressive reduction in $\Delta V_d$ at slower sweep rates, as shown in Fig.~\ref{Fig2}e, suggesting relaxation of the ferroelectric domains towards equilibrium.

\begin{figure*}
\includegraphics[width=1\textwidth]{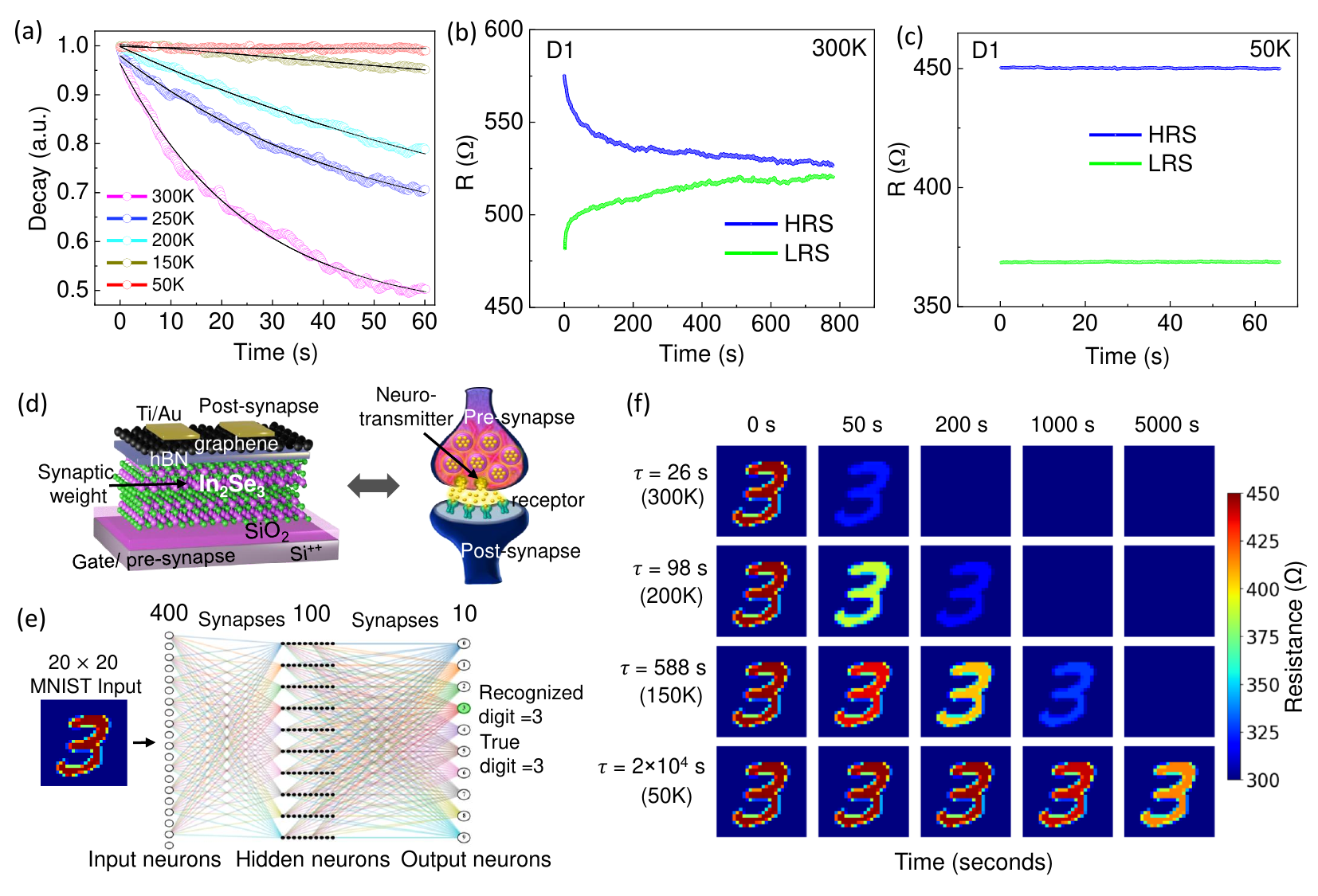}
\caption{(a) Post-pulse relaxation of normalized resistance at different temperatures for D1. (b, c) Retention stability of the high resistance state (HRS) and low resistance state (LRS) of device D1 at 300~K and 50~K, respectively. (d) Schematic illustration of the $\alpha$-In$_2$Se$_3$ based synaptic device and its analogy to biological neural function (e) Schematic illustration of the MNIST handwritten digit recognition procedure using the programmed conductance states of the  $\alpha$-In$_2$Se$_3$ based synaptic device (f) Temperature-dependent MNIST handwritten digit recognition using device D1, showing the evolution of recognition probability over time as a function of conductance state decay. The enhanced retention at 50~K preserves the programmed conductance states, enabling stable and robust recognition.  } \label{Fig3}
\end{figure*}

The lifetime of electrically written ferroelectric states is governed by the interplay of thermal fluctuations, polarization relaxation, and electrostatic boundary conditions. Hence, strong retention at room temperature is a critical requirement for nonvolatile memory and neuromorphic technologies. The temperature-dependent transport in the Graphene Fe-FET devices demonstrated a reduction in the $\Delta V_d$ values with increase in temperature as seen in the bottom panel of Fig.~\ref{Fig2}c. For device D1, the maximum polarization-induced charge density, $\Delta \sigma_p$ increases from $0.008 ~\mu C/cm^2$ to $0.028~\mu C/cm^2$ as the temperature is reduced from 300~K to 50~K, corresponding to an approximately four-fold enhancement at low temperature. A similar temperature behaviour was consistently observed in devices, D2 and D3, with D2 exhibiting the largest $\Delta V_d$ and consequently the highest polarization-induced charge density of $0.13~\mu C/cm^2$ below 100~K. (See supporting information)  To elucidate the origin of the temperature-dependent polarization response, potentiation and depression measurements were performed using controlled voltage pulse protocols. For potentiation, a constant gate pulse amplitude of -5~V with a width of 200~ms was applied, while number of pulses were incremented in steps of 10. Fig.~\ref{Fig2}f shows the corresponding evolution of graphene channel resistance at 300~K, 250~K, 150~K and 50~K, highlighting the stability of the resistance states at different temperatures. At 300~K, the resistance states display considerable relaxation immediately after the removal of the programming pulses. When cooled down to below 150~K, the resistance states relax considerably slower for the same pulse protocol, indicating an increased stability of the electrically written polarization states at cryogenic conditions. These resistance relaxation curves were normalized to the post-pulse resistance values and quantitatively analyzed by fitting with an exponential function $R(t)=R_0+ A\exp{(-t/\tau)}$ as shown in Fig.~\ref{Fig3}a, where $\tau$ is the characteristic relaxation time, while $R_0$ and $A$ are the fitting parameters. The extracted $\tau$ values increase systematically with decreasing temperature. At 300~K, $\tau\approx$ 26s~, while it increases remarkably to 588~s at 150~K signifying improved stability of polarization states. Below 100~K, the resistance states remain nearly saturated, with approximately 99.9\% retention ($\tau$ $\sim$ $2\times10^4$ s) over the measurement window. A direct comparison of the polarization retention stability at temperatures of 300~K and 50~K is shown in Fig.~\ref{Fig3}b \& c  respectively, for both the high‑resistance (HRS) and low‑resistance (LRS) states. The systematic increase in retention time at low temperatures suggests the key role of thermally activated ferroelectric domain dynamics in the relaxation of polarization-induced resistance states. The progressive change in resistance with increasing pulse number as seen in Fig.~\ref{Fig2}f is also consistent with such a cumulative ferroelectric domain reconfiguration, rather than abrupt charge trapping events. 

To further evaluate the neuromorphic functionality of device D1, MNIST handwritten-digit recognition was performed using the experimentally obtained potentiation-depression conductance states. \cite{Mohta2021,Kang2024,Shin2025,Wu2025,Song2024} Fig.~\ref{Fig3}d illustrates the analogy between biological synaptic signaling and the operation of the $\alpha$-In$_2$Se$_3$-based ferroelectric synaptic device. In biological neurons, neurotransmitter-mediated signaling regulates the postsynaptic response, whereas in the Graphene Fe-FET, electrically controlled polarization in $\alpha$-In$_2$Se$_3$ modulates the carrier density in the graphene channel, enabling programmable conductance states acting as synaptic weights. Fig.~\ref{Fig3}e illustrates the artificial neural network architecture employed for handwritten digit recognition using the MNIST dataset. Each $28\times28$ pixel image was downsampled to a $20\times20$ array, providing 400 input neurons connected to a hidden layer of 100 neurons and an output layer of 10 neurons corresponding to the digits 0-9. The synaptic weights were assigned using conductance values derived from the experimentally measured potentiation and depression characteristics of the device D1. During inference, the network processed the input pixel intensities through weighted connections and generated output activations for digit classification. The neuron with the highest output activation was assigned as the predicted digit, enabling evaluation of the device-derived conductance states for neuromorphic pattern recognition. The influence of state volatility on pattern recognition performance was examined from the evolution of MNIST recognition probability evaluated using 20 conductance states of D1 at different temperatures, as shown in Fig.~\ref{Fig3}f. At elevated temperatures, the programmed conductance states exhibited pronounced temporal relaxation, resulting in a progressive degradation of the corresponding synaptic weights and a consequent rapid decrease in the MNIST recognition probability over time. In contrast, at lower temperatures, where the conductance states were substantially more stable, the recognition probability was retained for a significantly longer duration of 5000~s and exhibited improved recognition stability at 50~K, consistent with the enhanced retention of the programmed resistance states at low temperatures. These results demonstrate that the temporal stability of multilevel conductance states is a critical parameter for reliable pattern recognition in ferroelectric synaptic devices.

\begin{figure*}
\includegraphics[width=1\textwidth]{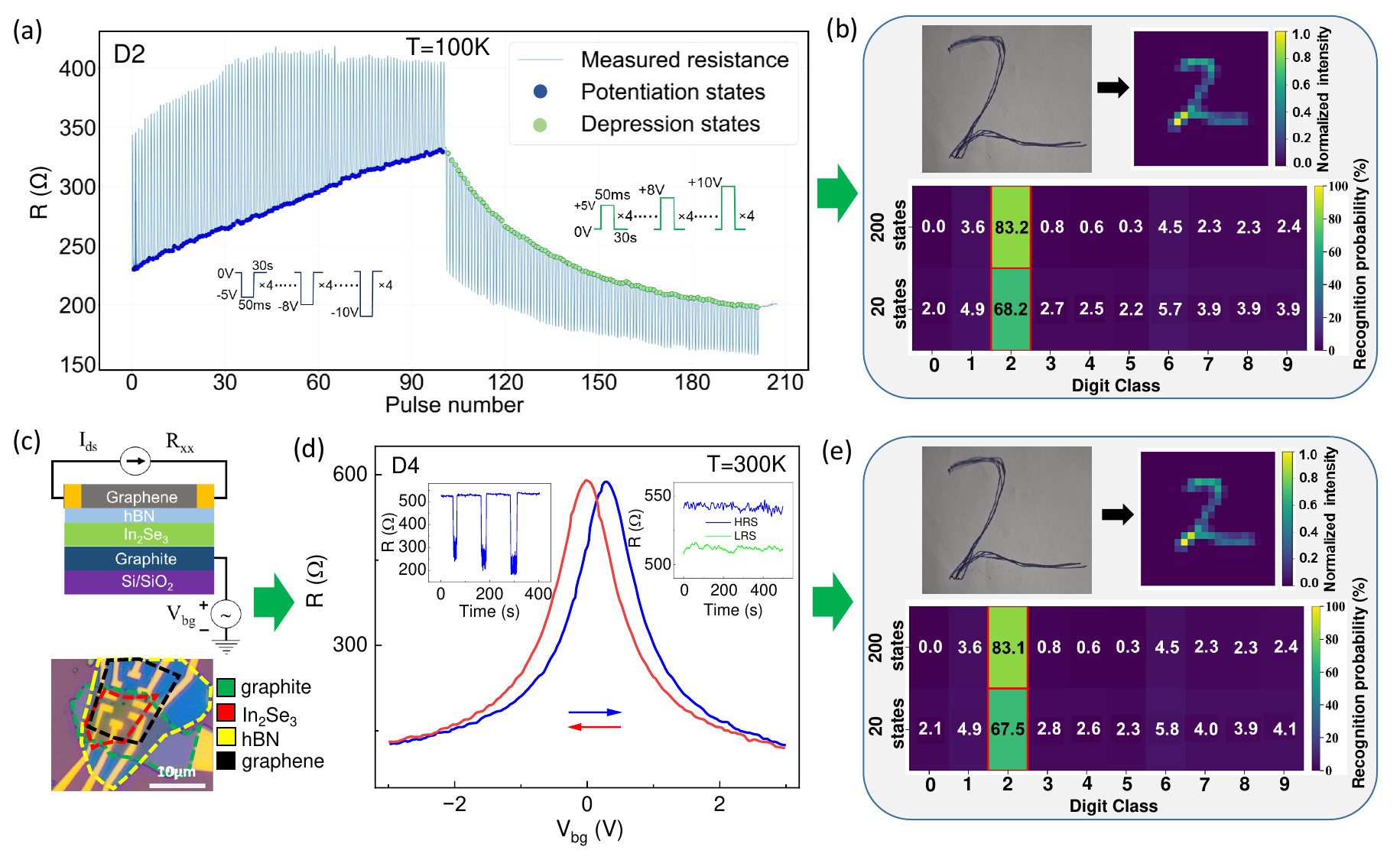}
\caption{(a) Ferroelectric synaptic behavior for device D2, demonstrating potentiation and depression across 200 different conductance states under a crescent pulse scheme. (b) Handwritten MNIST digit recognition accuracy using 200 and 20 conductance states of device D2 at 100~K.  (c) Schematic illustration and optical micrograph of the graphite-gated device D4 (scale bar $10\ \mu\mathrm{m}$) (d) R-V$_{bg}$ characteristics of device D4 showing clear Dirac point separation at 300~K. Left inset: Stable potentiation states of D4 at 300~K. Right inset: Stable HRS and LRS showing minimal resistance decay over the experimental time window. (e) Handwritten MNIST digit recognition accuracy using 200 and 20 conductance states of device D4. }  \label{Fig4}
\end{figure*}

To assess the neuromorphic potential of Graphene Fe-FETs, we next evaluated the recognition of handwritten-digits using conductance states extracted from the transfer characteristics of device D2. The MNIST dataset, comprising $28\times28$ pixel grayscale images of handwritten digits (0-9), was used to assess the impact of the number of accessible conductance states on recognition performance. The experimentally obtained resistance states were converted into conductance values and used to represent the synaptic weights in the recognition model. The recognition performance was evaluated using two conductance-state configurations, comprising 200 and 20 states, to investigate the effect of conductance resolution on classification accuracy. To mitigate room-temperature resistance state relaxation, stable 200 conductance states were generated in D2 at 100~K by programming crescent pulses of $\pm$~5 to $\pm$~10~V with a width of 50~ms and a 30~s interval as shown in Fig.~\ref{Fig4}a. Using these experimentally derived conductance levels, the MNIST recognition model achieved calibration probabilities exceeding 99\% for individual digit classes, while the recognition probability for additional handwritten samples of the corresponding digits exceeded 80\% as can be seen from Fig.~\ref{Fig4}b. When the number of conductance states was reduced to 20 using an alternative pulse scheme, the recognition probability decreased to 68\%, highlighting the influence of conductance state resolution on pattern recognition performance.

Next we explore the role of surface-charge-induced depolarization fields of $\alpha$-In$_2$Se$_3$ in governing the ferroelectric state stability in Graphene Fe-FETs. In an unscreened or weakly screened $\alpha$-In$_2$Se$_3$ layer, the out-of-plane polarization produces bound surface charges \cite{Huang2024}. In the absence of sufficient compensating free charge, these bound charges generate a depolarization field that opposes the polarization.\cite{Wurfel1973,Nahid2024,Shuai2023} The resulting electrostatic instability promotes relaxation towards a lower energy polarization configuration. For device D1, the ferroelectric surface is in contact with a dielectric, and the depolarization field can be expressed as 
\cite{Watanabe2021}

\begin{equation}    
 \lvert E_{dep} \rvert = \frac{P_{\perp}}{\epsilon_0 . \epsilon_{FE}} \frac{1}{[1 + \frac{d_{FE}}{\epsilon_{FE}} / (\frac{\epsilon_{hBN}}{d_{hBN}} + \frac{\epsilon_{SiO_2}}{d_{SiO_2}})]}
\end{equation}

where $P_{\perp}$ is the out-of-plane polarization in $\alpha$-In$_2$Se$_3$, $\epsilon$ is the dielectric constant and  $d$ is the thickness of the dielectric layers.  Using the theoretically predicted value of $P_{\perp}$ = 0.9 $\mu C/cm^2$, \cite{Ding2017, Wan2019} and $\epsilon_{FE}$ = 17 for In$_2$Se$_3$ of 20~nm thickness,\cite{Wu2015} the numerical value of the depolarization field is calculated to be 592~kV/cm. The thermal fluctuations aid the depolarization field in the reorientation of polarization domains, leading to relaxation of the resistance state. Consequently, cooling increases the stability of polarization configuration by reducing thermal activation of domain instabilities. Hence the retention characteristics are governed by two competing energy scales; the electrostatic energy associated with the depolarization field and the thermal energy available for polarization fluctuations. At room temperature both effects contribute towards the rapid relaxation of the programmed resistance states.  Upon cooling, the reduction in thermal energy progressively suppresses polarization relaxation, allowing the remnant polarization to persist despite the presence of the depolarization field. 

While the reduction of temperature effectively suppresses the thermal fluctuations, screening the polarization-induced bound charges provides a more technologically relevant route to room-temperature computing. The depolarization field can be mitigated to stabilize the polarization configuration by effective electrostatic screening using metallic gates, high-$\kappa$ dielectrics or optimized interface engineering. To directly probe the screening effects, a graphite electrode was introduced in direct contact with the In$_2$Se$_3$ layer as shown in Fig.~\ref{Fig4}c for device D4. Graphite provides a high-density conducting boundary that effectively screens the polarization-induced surface charge while preserving the vdW interface. From the Dirac point separation in D4 at room temperature, $\Delta V_d$, as shown in Fig.~\ref{Fig4}d, the corresponding polarization-induced charge density extracted is $\approx0.042~\mu C/cm^2$. This is comparable to the values obtained in D1-D3 at low temperatures. For a conducting graphite boundary, the residual depolarization field can be approximated as \cite{Mehta1973} 

\begin{equation}    
\lvert E_{dep} \rvert = \frac{P_{\perp}}{\varepsilon_0\epsilon_{FE}} \cdot \frac{\epsilon_{FE}/l}{\epsilon_{FE}/l+\epsilon_g/l_s}
\end{equation}

where $l_s$ is the screening length of the graphite electrode and $l$ is the thickness of $\alpha$-In$_2$Se$_3$. Taking  $\epsilon_g$ = 3$\epsilon_0$, the out-of-plane dielectric constant for multilayer graphene \cite{Santos2013}, $l_s \approx 0.4$~nm \cite{Divincenzo1984} and $l\approx$ 20~nm, the screened depolarization field was estimated to be $\approx$ 221~kV/cm. Thus, graphite electrode reduces the depolarization field from approximately 592~kV/cm to 221~kV/cm, corresponding to a reduction of roughly 66\%.

The polarization switching characteristics of D4 further demonstrate the critical role of electrostatic screening. Application of  100 pulses of width 50~ms, with an amplitude of $\pm$~3~V produced well-defined resistance states, and, importantly, these states remained considerably stable after the removal of the programming voltage as shown in the left inset of Fig.~ \ref{Fig4}d. Highly stable HRS and LRS states were observed as shown in the right inset of Fig.~\ref{Fig4}d, indicating that the robust nature of programmed states post removal of voltage pulse. The contrasting behavior between D1-D3 and D4 at 300~K provides strong evidence that the retention of the polarization states is governed by the electrostatic boundary conditions. The substantial reduction in depolarization field in D4 leads to stable resistance states even at room temperature, where despite considerable thermal fluctuations, only a small decay rate of order $\sim 10^4$~s is observed. The graphite-gated device D4 exhibited stable multilevel resistance states at room temperature, enabling the conductance states required for MNIST recognition. The performance for both the 200 and 20-state configurations was comparable to that obtained for D2 (at 100~K) as shown in Fig.~\ref{Fig4}e. This successfully demonstrates multilevel conductance modulation that can be used for handwritten-digit recognition at room temperature, underscoring the need for screening the depolarization field in 2D ferroelectrics to achieve reliable neuromorphic functionality in ferroelectric heterostructure devices.

Overall, these observations establish a unified picture in which the retention of  graphene resistance states is controlled by the interplay between ferroelectric polarization, depolarization field, and thermal fluctuations. Weak electrostatic screening results in unstable room temperature polarization states, while cooling suppresses thermal fluctuations and progressively enhances the remanent polarization-induced charge density. Alternatively, introducing a metallic graphite boundary efficiently screens the polarization-induced bound charges and suppresses the depolarization field, enabling robust resistance states at room temperature. Thus, electrostatic boundary engineering delivers an efficient route to achieve room-temperature nonvolatile resistance states in $\alpha$-In$_2$Se$_3$/graphene heterostructures.

\begin{acknowledgement}
P.G., S.P., V.P., A.M.B., and V.K. acknowledge funding support from the DST- Nanomission program of the Department of Science and Technology, Government of India (DST/NM/TUE/QM-1/2019). This material is based upon work supported by the Air Force Office of Scientific Research under award numbers FA2386-24-1-4093 and FA2386-26-1-4002. P.C. and A.N.P. acknowledge the Thematic Unit of Excellence on Nanodevice Technology (Grant No. SR/NM/NS-09/2011) and the Technical Research Centre (TRC) Instrument facilities of S. N. Bose National Centre for Basic Sciences, established under the TRC project of Department of Science and Technology (DST), Govt. of India. A.N.P. acknowledges DST Nano Mission: DST/NM/TUE/QM-10/2019. K.P.M thanks the Anusandhan National Research Foundation (ANRF), India, for financial support under the National Post-Doctoral Fellowship (NPDF) scheme [File Number:PDF/2023/001249]. K.W. and T.T. acknowledge support from the the CREST (JPMJCR24A5), JST and World Premier International Research Center Initiative (WPI), MEXT, Japan.

\end{acknowledgement}

\bibliography{Reference}

@article{Raman_In2Se3_MRB,
title = {Raman scattering in $\alpha$-In$_2$Se$_3$ crystals},
journal = {Materials Research Bulletin},
volume = {36},
number = {15},
pages = {2577-2583},
year = {2001},
issn = {0025-5408},
doi = {https://doi.org/10.1016/S0025-5408(01)00746-2},
url = {https://www.sciencedirect.com/science/article/pii/S0025540801007462},
author = {R. Lewandowska and R. Bacewicz and J. Filipowicz and W. Paszkowicz}
}

@article{Chang2016,
   author = {Kai Chang and Junwei Liu and Haicheng Lin and Na Wang and Kun Zhao and Anmin Zhang and Feng Jin and Yong Zhong and Xiaopeng Hu and Wenhui Duan and Qingming Zhang and Liang Fu and Qi-Kun Xue and Xi Chen and Shuai-Hua Ji},
   doi = {10.1126/science.aad8609},
   issn = {0036-8075},
   issue = {6296},
   journal = {Science},
   month = {7},
   pages = {274-278},
   title = {Discovery of robust in-plane ferroelectricity in atomic-thick SnTe},
   volume = {353},
   year = {2016}
}

@article{Liu2016,
   author = {Fucai Liu and Lu You and Kyle L. Seyler and Xiaobao Li and Peng Yu and Junhao Lin and Xuewen Wang and Jiadong Zhou and Hong Wang and Haiyong He and Sokrates T. Pantelides and Wu Zhou and Pradeep Sharma and Xiaodong Xu and Pulickel M. Ajayan and Junling Wang and Zheng Liu},
   doi = {10.1038/ncomms12357},
   issn = {20411723},
   journal = {Nature Communications},
   month = {8},
   pages = {12357},
   publisher = {Nature Publishing Group},
   title = {Room-temperature ferroelectricity in CuInP$_2$S$_6$ ultrathin flakes},
   volume = {7},
   year = {2016}
}

@article{Ding2017,
   author = {Wenjun Ding and Jianbao Zhu and Zhe Wang and Yanfei Gao and Di Xiao and Yi Gu and Zhenyu Zhang and Wenguang Zhu},
   doi = {10.1038/ncomms14956},
   issn = {2041-1723},
   journal = {Nature Communications },
   month = {4},
   pages = {14956},
   pmid = {28387225},
   publisher = {Nature Publishing Group},
   title = {Prediction of intrinsic two-dimensional ferroelectrics in In$_2$Se$_3$ and other III$_2$-VI$_3$ van der Waals materials},
   volume = {8},
   year = {2017}
}

@article{Xue2019,
   doi = {10.1002/adma.201901300},
   issn = {15214095},
   issue = {29},
   journal = {Advanced Materials},
   month = {7},
   pmid = {31148294},
   pages = {1901300},
   publisher = {Wiley-VCH Verlag},
   title = {Gate-Tunable and Multidirection-Switchable Memristive Phenomena in a Van Der Waals Ferroelectric},
   volume = {31},
   year = {2019}
}

@article{Zhou2017,
   author = {Yu Zhou and Di Wu and Yihan Zhu and Yujin Cho and Qing He and Xiao Yang and Kevin Herrera and Zhaodong Chu and Yu Han and Michael C. Downer and Hailin Peng and Keji Lai},
   doi = {10.1021/acs.nanolett.7b02198},
   issn = {15306992},
   issue = {9},
   journal = {Nano Letters},
   month = {9},
   pages = {5508-5513},
   pmid = {28841328},
   publisher = {American Chemical Society},
   title = {Out-of-Plane Piezoelectricity and Ferroelectricity in Layered $\alpha$-In$_2$Se$_3$ Nanoflakes},
   volume = {17},
   year = {2017}
}

@article{Cui2018,
   author = {Chaojie Cui and Wei-Jin Hu and Xingxu Yan and Christopher Addiego and Wenpei Gao and Yao Wang and Zhe Wang and Linze Li and Yingchun Cheng and Peng Li and Xixiang Zhang and Husam N. Alshareef and Tom Wu and Wenguang Zhu and Xiaoqing Pan and Lain-Jong Li},
   doi = {10.1021/acs.nanolett.7b04852},
   issn = {1530-6984},
   issue = {2},
   journal = {Nano Letters},
   month = {2},
   pages = {1253-1258},
   title = {Intercorrelated In-Plane and Out-of-Plane Ferroelectricity in Ultrathin Two-Dimensional Layered Semiconductor In$_2$Se$_3$},
   volume = {18},
   year = {2018}
}

@article{Liu2019,
   author = {Lixuan Liu and Jiyu Dong and Junquan Huang and Anmin Nie and Kun Zhai and Jianyong Xiang and Bochong Wang and Fusheng Wen and Congpu Mu and Zhisheng Zhao and Yongji Gong and Yongjun Tian and Zhongyuan Liu},
   doi = {10.1021/acs.chemmater.9b03499},
   issn = {15205002},
   issue = {24},
   journal = {Chemistry of Materials},
   month = {12},
   pages = {10143-10149},
   publisher = {American Chemical Society},
   title = {Atomically Resolving Polymorphs and Crystal Structures of In$_2$Se$_3$},
   volume = {31},
   year = {2019}
}

@article{Xu2020,
   author = {Kai Xu and Wei Jiang and Xueshi Gao and Zijing Zhao and Tony Low and Wenjuan Zhu},
   doi = {10.1039/d0nr06872a},
   issn = {20403372},
   issue = {46},
   journal = {Nanoscale},
   month = {12},
   pages = {23488-23496},
   pmid = {33211783},
   publisher = {Royal Society of Chemistry},
   title = {Optical control of ferroelectric switching and multifunctional devices based on van der Waals ferroelectric semiconductors},
   volume = {12},
   year = {2020}
}

@article{Dutta2021,
   author = {Debopriya Dutta and Subhrajit Mukherjee and Michael Uzhansky and Elad Koren},
   doi = {10.1038/s41699-021-00261-w},
   issn = {2397-7132},
   issue = {1},
   journal = {npj 2D Materials and Applications},
   month = {9},
   pages = {81},
   title = {Cross-field optoelectronic modulation via inter-coupled ferroelectricity in 2D In$_2$Se$_3$},
   volume = {5},
   year = {2021}
}

@article{Guo2025,
   author = {Zhenzhou Guo and Shifeng Qian and Xiaodong Zhou and Wenhong Wang and Zhenxiang Cheng and Xiaotian Wang},
   doi = {10.1038/s41467-025-67240-3},
   issn = {2041-1723},
   issue = {1},
   journal = {Nature Communications},
   month = {12},
   pages = {549},
   title = {Sliding ferroelectric metal with ferrimagnetism},
   volume = {17},
   year = {2025}
}

@article{Wan2019,
   author = {Siyuan Wan and Yue Li and Wei Li and Xiaoyu Mao and Chen Wang and Chen Chen and Jiyu Dong and Anmin Nie and Jianyong Xiang and Zhongyuan Liu and Wenguang Zhu and Hualing Zeng},
   doi = {10.1002/adfm.201808606},
   issn = {16163028},
   issue = {20},
   journal = {Advanced Functional Materials},
   month = {5},
   pages = {1808606},
   publisher = {Wiley-VCH Verlag},
   title = {Nonvolatile Ferroelectric Memory Effect in Ultrathin $\alpha$-In$_2$Se$_3$},
   volume = {29},
   year = {2019}
}

@article{Sahota2004,
   author = {H. Sahota},
   doi = {10.1007/s00161-003-0148-y},
   issn = {09351175},
   issue = {1-2},
   journal = {Continuum Mechanics and Thermodynamics},
   pages = {163-175},
   publisher = {Springer New York},
   title = {Simulation of butterfly loops in ferroelectric materials},
   volume = {16},
   year = {2004}
}

@article{Watanabe2021,
   author = {Yukio Watanabe},
   doi = {10.1038/s41598-021-81237-0},
   issn = {2045-2322},
   issue = {1},
   journal = {Scientific Reports},
   month = {1},
   pages = {2155},
   title = {Examination of permittivity for depolarization field of ferroelectric by ab initio calculation, suggesting hidden mechanisms},
   volume = {11},
   year = {2021}
}

@article{Wu2015,
   author = {Di Wu and Alexander J. Pak and Yingnan Liu and Yu Zhou and Xiaoyu Wu and Yihan Zhu and Min Lin and Yu Han and Yuan Ren and Hailin Peng and Yu Hao Tsai and Gyeong S. Hwang and Keji Lai},
   doi = {10.1021/acs.nanolett.5b03575},
   issn = {15306992},
   issue = {12},
   journal = {Nano Letters},
   month = {12},
   pages = {8136-8140},
   publisher = {American Chemical Society},
   title = {Thickness-Dependent Dielectric Constant of Few-Layer In$_2$Se$_3$ Nanoflakes},
   volume = {15},
   year = {2015}
}

@article{Huang2024,
   author = {Jiawei Huang and Changming Ke and Zhuang Qian and Shi Liu},
   doi = {10.1021/acs.nanolett.4c01362},
   issn = {15306992},
   issue = {22},
   journal = {Nano Letters},
   month = {6},
   pages = {6683-6688},
   pmid = {38767925},
   publisher = {American Chemical Society},
   title = {Competing Charge Transfer and Screening Effects in Two-Dimensional Ferroelectric Capacitors},
   volume = {24},
   year = {2024}
}

@article{Nahid2024,
   author = {Shahriar Muhammad Nahid and Sung Woo Nam and Arend M. van der Zande},
   doi = {10.1021/acsnano.3c11558},
   issn = {1936086X},
   issue = {22},
   journal = {ACS Nano},
   month = {6},
   pages = {14198-14206},
   pmid = {38771928},
   publisher = {American Chemical Society},
   title = {Depolarization Field-Induced Photovoltaic Effect in Graphene/$\alpha$-In$_2$Se$_3$/Graphene Heterostructures},
   volume = {18},
   year = {2024}
}

@article{Divincenzo1984,
  title = {Self-consistent effective-mass theory for intralayer screening in graphite intercalation compounds},
  author = {DiVincenzo, D. P. and Mele, E. J.},
  journal = {Phys. Rev. B},
  volume = {29},
  issue = {4},
  pages = {1685--1694},
  numpages = {0},
  year = {1984},
  month = {Feb},
  publisher = {American Physical Society},
  doi = {10.1103/PhysRevB.29.1685},
  url = {https://link.aps.org/doi/10.1103/PhysRevB.29.1685}
}

@article{Santos2013,
   author = {Elton J.G. Santos and Efthimios Kaxiras},
   doi = {10.1021/nl303611v},
   issn = {15306984},
   issue = {3},
   journal = {Nano Letters},
   month = {3},
   pages = {898-902},
   title = {Electric-field dependence of the effective dielectric constant in graphene},
   volume = {13},
   year = {2013}
}

@article{Mehta1973,
   author = {R. R. Mehta and B. D. Silverman and J. T. Jacobs},
   doi = {10.1063/1.1662770},
   issn = {00218979},
   issue = {8},
   journal = {Journal of Applied Physics},
   pages = {3379-3385},
   title = {Depolarization fields in thin ferroelectric films},
   volume = {44},
   year = {1973}
}

@article{Huang2022,
   author = {Yu Ting Huang and Nian Ke Chen and Zhen Ze Li and Xue Peng Wang and Hong Bo Sun and Shengbai Zhang and Xian Bin Li},
   doi = {10.1002/inf2.12341},
   issn = {25673165},
   issue = {8},
   journal = {InfoMat},
   pages = {e12341},
   month = {8},
   title = {Two-dimensional In$_2$Se$_3$: A rising advanced material for ferroelectric data storage},
   volume = {4},
   year = {2022}
}

@article{Wang2020,
author = {Wang, Lin and Wang, Xiaojie and Zhang, Yishu and Li, Runlai and Ma, Teng and Leng, Kai and Chen, Zhi and Abdelwahab, Ibrahim and Loh, Kian Ping},
title = {Exploring Ferroelectric Switching in $\alpha$-In$_2$Se$_3$ for Neuromorphic Computing},
journal = {Advanced Functional Materials},
volume = {30},
number = {45},
pages = {2004609},
doi = {https://doi.org/10.1002/adfm.202004609},
url = {https://advanced.onlinelibrary.wiley.com/doi/abs/10.1002/adfm.202004609},
eprint = {https://advanced.onlinelibrary.wiley.com/doi/pdf/10.1002/adfm.202004609},
year = {2020}
}

@article{Zhang2022,
   author = {Dawei Zhang and Peggy Schoenherr and Pankaj Sharma and Jan Seidel},
   doi = {10.1038/s41578-022-00484-3},
   issn = {2058-8437},
   issue = {1},
   journal = {Nature Reviews Materials},
   month = {10},
   pages = {25-40},
   title = {Ferroelectric order in van der Waals layered materials},
   volume = {8},
   year = {2022}
}

@article{Wang2023,
   author = {Chuanshou Wang and Lu You and David Cobden and Junling Wang},
   doi = {10.1038/s41563-022-01422-y},
   issn = {1476-1122},
   issue = {5},
   journal = {Nature Materials},
   month = {5},
   pages = {542-552},
   title = {Towards two-dimensional van der Waals ferroelectrics},
   volume = {22},
   year = {2023}
}

@article{Wang2025,
   author = {Hao Wang and Yao Wen and Hui Zeng and Ziren Xiong and Yangyuan Tu and Hao Zhu and Ruiqing Cheng and Lei Yin and Jian Jiang and Baoxing Zhai and Chuansheng Liu and Chongxin Shan and Jun He},
   doi = {10.1002/adma.202305044},
   issn = {0935-9648},
   issue = {31},
   journal = {Advanced Materials},
   month = {8},
   pages = {2305044},
   title = {2D Ferroic Materials for Nonvolatile Memory Applications},
   volume = {37},
   year = {2025}
}

@article{Liu2021,
   author = {Minghao Liu and Ting Liao and Ziqi Sun and Yuantong Gu and Liangzhi Kou},
   doi = {10.1039/D1CP02788C},
   issn = {1463-9076},
   issue = {38},
   journal = {Physical Chemistry Chemical Physics},
   pages = {21376-21384},
   title = {2D ferroelectric devices: working principles and research progress},
   volume = {23},
   year = {2021}
}

@article{Li2018screening,
   author = {Mingqiang Li and Xiaoxing Cheng and Ning Li and Heng-Jui Liu and Yen-Lin Huang and Kaihui Liu and Ying-Hao Chu and Dapeng Yu and Long-Qing Chen and Yuichi Ikuhara and Peng Gao},
   doi = {10.1103/PhysRevB.97.180103},
   issn = {2469-9950},
   issue = {18},
   journal = {Physical Review B},
   month = {5},
   pages = {180103},
   title = {Atomic-scale mechanism of internal structural relaxation screening at polar interfaces},
   volume = {97},
   year = {2018}
}

@article{Junquera2003,
   author = {Javier Junquera and Philippe Ghosez},
   doi = {10.1038/nature01501},
   issn = {0028-0836},
   issue = {6931},
   journal = {Nature},
   month = {4},
   pages = {506-509},
   title = {Critical thickness for ferroelectricity in perovskite ultrathin films},
   volume = {422},
   year = {2003}
}

@article{Gao2021,
   author = {Jing Gao and Yue Zheng and Wei Yu and Yanan Wang and Tengyu Jin and Xuan Pan and Kian Ping Loh and Wei Chen},
   doi = {10.1002/smm2.1020},
   issn = {2688-819X},
   issue = {1},
   journal = {SmartMat},
   month = {3},
   pages = {88-98},
   title = {Intrinsic polarization coupling in 2D $\alpha$-In$_2$Se$_3$ toward artificial synapse with multimode operations},
   volume = {2},
   year = {2021}
}

@article{Li2020,
   author = {Yue Li and Chen Chen and Wei Li and Xiaoyu Mao and Heng Liu and Jianyong Xiang and Anmin Nie and Zhongyuan Liu and Wenguang Zhu and Hualing Zeng},
   doi = {10.1002/aelm.202000061},
   issn = {2199-160X},
   issue = {7},
   journal = {Advanced Electronic Materials},
   month = {7},
   pages = {2000061},
   title = {Orthogonal Electric Control of the Out‐Of‐Plane Field‐Effect in 2D Ferroelectric $\alpha$-In$_2$Se$_3$},
   volume = {6},
   year = {2020}
}

@article{Ghani2024,
   author = {Maheera Abdul Ghani and Soumya Sarkar and Yang Li and Ye Wang and Kenji Watanabe and Takashi Taniguchi and Yan Wang and Manish Chhowalla},
   doi = {10.1557/s43581-024-00109-y},
   issn = {2329-2237},
   issue = {2},
   journal = {MRS Energy \& Sustainability},
   month = {8},
   pages = {616-623},
   title = {Ferroelectric field effect transistors based on two-dimensional CuInP$_2$S$_6$ (CIPS) and graphene heterostructures},
   volume = {11},
   year = {2024}
}

@article{Paul2026,
   author = {Sourav Paul and Prasenjit Ghosh and Krishna Prasad Maity and Vineet Pandey and Abhijith M. B. and Premananda Chatterjee and Rinku Paul and Kenji Watanabe and Takashi Taniguchi and Nicholas R. Glavin and Ajit K. Roy and Atindra Nath Pal and Vidya Kochat},
   doi = {10.1039/D6NR01832G},
   issn = {2040-3364},
   issue = {34},
   journal = {Nanoscale},
   pages = {18299-18307},
   title = {Impact of disorder dynamics and multi-domain kinetics on the sliding ferroelectricity of CVD-grown 3R-WSe$_2$ bilayers},
   volume = {18},
   year = {2026}
}

@article{Dean2010,
   author = {C. R. Dean and A. F. Young and I. Meric and C. Lee and L. Wang and S. Sorgenfrei and K. Watanabe and T. Taniguchi and P. Kim and K. L. Shepard and J. Hone},
   doi = {10.1038/nnano.2010.172},
   issn = {1748-3387},
   issue = {10},
   journal = {Nature Nanotechnology},
   month = {10},
   pages = {722-726},
   title = {Boron nitride substrates for high-quality graphene electronics},
   volume = {5},
   year = {2010}
}

@article{Park2015,
   author = {Nahee Park and Haeyong Kang and Jeongmin Park and Yourack Lee and Yoojoo Yun and Jeong-Ho Lee and Sang-Goo Lee and Young Hee Lee and Dongseok Suh},
   doi = {10.1021/acsnano.5b04339},
   issn = {1936-0851},
   issue = {11},
   journal = {ACS Nano},
   month = {11},
   pages = {10729-10736},
   title = {Ferroelectric Single-Crystal Gated Graphene/Hexagonal-BN/Ferroelectric Field-Effect Transistor},
   volume = {9},
   year = {2015}
}

@article{Mohta2021,
   author = {Neha Mohta and Ankit Rao and Nayana Remesh and R. Muralidharan and Digbijoy N. Nath},
   doi = {10.1039/D1RA07728G},
   issn = {2046-2069},
   issue = {58},
   journal = {RSC Advances},
   pages = {36901-36912},
   title = {An artificial synaptic transistor using an $\alpha$-In$_2$Se$_3$ van der Waals ferroelectric channel for pattern recognition},
   volume = {11},
   year = {2021}
}

@article{Kang2024,
   author = {Seok‐Ju Kang and Wonzee Jung and Oh Hun Gwon and Han Seul Kim and Hye Ryung Byun and Jong Yun Kim and Seo Gyun Jang and BeomKyu Shin and Ojun Kwon and Byungjin Cho and Kanghoon Yim and Young‐Jun Yu},
   doi = {10.1002/smll.202307346},
   issn = {1613-6810},
   issue = {22},
   journal = {Small},
   pages = {2307346},
   month = {5},
   title = {Photo‐Assisted Ferroelectric Domain Control for $\alpha$-In$_2$Se$_3$ Artificial Synapses Inspired by Spontaneous Internal Electric Fields},
   volume = {20},
   year = {2024}
}

@article{Song2024,
   author = {Chong‐Myeong Song and Dongha Kim and Shinbuhm Lee and Hyuk‐Jun Kwon},
   doi = {10.1002/advs.202308588},
   issn = {2198-3844},
   issue = {16},
   journal = {Advanced Science},
   month = {4},
   pages = {2308588},
   title = {Ferroelectric 2D SnS$_2$ Analog Synaptic FET},
   volume = {11},
   year = {2024}
}

@article{Shin2025,
   author = {Jaeho Shin and Jingon Jang and Chi Hun Choi and Jaegyu Kim and Lucas Eddy and Phelecia Scotland and Lane W. Martin and Yimo Han and James M. Tour},
   doi = {10.1002/aelm.202400603},
   issn = {2199-160X},
   issue = {5},
   journal = {Advanced Electronic Materials},
   month = {4},
   pages = {2400603},
   title = {In$_2$Se$_3$ Synthesized by the FWF Method for Neuromorphic Computing},
   volume = {11},
   year = {2025}
}

@article{Wu2025,
   author = {Yuxuan Wu and Ning Jiang and Xiaojun Qiao and Wenping Geng and Xiujian Chou},
   doi = {10.1021/acsami.5c18166},
   issn = {1944-8244},
   issue = {47},
   journal = {ACS Applied Materials \& Interfaces},
   month = {11},
   pages = {64921-64930},
   title = {Two-Terminal Ferroelectric Artificial Synaptic Devices with Asymmetric Structure},
   volume = {17},
   year = {2025}
}

@article{Shuai2023,
   author = {Wen-Jun Shuai and Rui Wang and Jin-Zhu Zhao},
   doi = {10.1103/PhysRevB.107.155427},
   issn = {2469-9950},
   issue = {15},
   journal = {Physical Review B},
   month = {4},
   pages = {155427},
   title = {Ferroelectric phase transition driven by anharmonic lattice mode coupling in two-dimensional monolayer In$_2$Se$_3$},
   volume = {107},
   year = {2023}
}

@article{Wurfel1973,
   author = {P. Wurfel and I. P. Batra},
   doi = {10.1103/PhysRevB.8.5126},
   issn = {0556-2805},
   issue = {11},
   journal = {Physical Review B},
   month = {12},
   pages = {5126-5133},
   title = {Depolarization-Field-Induced Instability in Thin Ferroelectric Films—Experiment and Theory},
   volume = {8},
   year = {1973}
}

@article{Scott2007,
  title = {Applications of Modern Ferroelectrics},
  volume = {315},
  ISSN = {1095-9203},
  url = {http://dx.doi.org/10.1126/science.1129564},
  DOI = {10.1126/science.1129564},
  number = {5814},
  journal = {Science},
  publisher = {American Association for the Advancement of Science (AAAS)},
  author = {Scott,  J. F.},
  year = {2007},
  month = Feb,
  pages = {954–959}
}

@article{Khan2020,
  title = {The future of ferroelectric field-effect transistor technology},
  volume = {3},
  ISSN = {2520-1131},
  url = {http://dx.doi.org/10.1038/s41928-020-00492-7},
  DOI = {10.1038/s41928-020-00492-7},
  number = {10},
  journal = {Nature Electronics},
  publisher = {Springer Science and Business Media LLC},
  author = {Khan,  Asif Islam and Keshavarzi,  Ali and Datta,  Suman},
  year = {2020},
  month = Oct,
  pages = {588–597}
}

@article{Xue2018,
  title = {Multidirection Piezoelectricity in Mono- and Multilayered Hexagonal $\alpha$-In$_2$Se$_3$},
  volume = {12},
  ISSN = {1936-086X},
  url = {http://dx.doi.org/10.1021/acsnano.8b02152},
  DOI = {10.1021/acsnano.8b02152},
  number = {5},
  journal = {ACS Nano},
  publisher = {American Chemical Society (ACS)},
  author = {Xue,  Fei and Zhang,  Junwei and Hu,  Weijin and Hsu,  Wei-Ting and Han,  Ali and Leung,  Siu-Fung and Huang,  Jing-Kai and Wan,  Yi and Liu,  Shuhai and Zhang,  Junli and He,  Jr-Hau and Chang,  Wen-Hao and Wang,  Zhong Lin and Zhang,  Xixiang and Li,  Lain-Jong},
  year = {2018},
  month = Apr,
  pages = {4976–4983}
}

@article{Xue2018_2,
  title = {Room‐Temperature Ferroelectricity in Hexagonally Layered $\alpha$-In$_2$Se$_3$ Nanoflakes down to the Monolayer Limit},
  volume = {28},
  ISSN = {1616-3028},
  url = {http://dx.doi.org/10.1002/adfm.201803738},
  DOI = {10.1002/adfm.201803738},
  number = {50},
  journal = {Advanced Functional Materials},
  publisher = {Wiley},
  author = {Xue,  Fei and Hu,  Weijin and Lee,  Ko‐Chun and Lu,  Li‐Syuan and Zhang,  Junwei and Tang,  Hao‐Ling and Han,  Ali and Hsu,  Wei‐Ting and Tu,  Shaobo and Chang,  Wen‐Hao and Lien,  Chen‐Hsin and He,  Jr‐Hau and Zhang,  Zhidong and Li,  Lain‐Jong and Zhang,  Xixiang},
  year = {2018},
  month = Oct ,
  pages = {1803738}
}

@article{Joo2023,
  title = {Memory and Synaptic Devices Based on Emerging 2D Ferroelectricity},
  volume = {9},
  ISSN = {2199-160X},
  url = {http://dx.doi.org/10.1002/aelm.202300211},
  DOI = {10.1002/aelm.202300211},
  number = {8},
  journal = {Advanced Electronic Materials},
  publisher = {Wiley},
  author = {Joo,  Yanggeun and Hwang,  Eunji and Hong,  Heemyoung and Cho,  Suyeon and Yang,  Heejun},
  year = {2023},
  month = June ,
  pages = {2300211}
}

@article{Huang2025,
  title = {Two-dimensional ferroelectric synaptic devices enabled by diverse coupling mechanisms},
  volume = {17},
  ISSN = {2040-3372},
  url = {http://dx.doi.org/10.1039/d5nr02639c},
  DOI = {10.1039/d5nr02639c},
  number = {39},
  journal = {Nanoscale},
  publisher = {Royal Society of Chemistry (RSC)},
  author = {Huang,  Youna and Wang,  Wei and Li,  Yang and Li,  Changjian},
  year = {2025},
  pages = {22763–22794}
}

@article{Yasuda2021,
  title = {Stacking-engineered ferroelectricity in bilayer boron nitride},
  volume = {372},
  ISSN = {1095-9203},
  url = {http://dx.doi.org/10.1126/science.abd3230},
  DOI = {10.1126/science.abd3230},
  number = {6549},
  journal = {Science},
  publisher = {American Association for the Advancement of Science (AAAS)},
  author = {Yasuda,  Kenji and Wang,  Xirui and Watanabe,  Kenji and Taniguchi,  Takashi and Jarillo-Herrero,  Pablo},
  year = {2021},
  month = June,
  pages = {1458–1462}
}

@article{Yasuda2024,
  title = {Ultrafast high-endurance memory based on sliding ferroelectrics},
  volume = {385},
  ISSN = {1095-9203},
  url = {http://dx.doi.org/10.1126/science.adp3575},
  DOI = {10.1126/science.adp3575},
  number = {6704},
  journal = {Science},
  publisher = {American Association for the Advancement of Science (AAAS)},
  author = {Yasuda,  Kenji and Zalys-Geller,  Evan and Wang,  Xirui and Bennett,  Daniel and Cheema,  Suraj S. and Watanabe,  Kenji and Taniguchi,  Takashi and Kaxiras,  Efthimios and Jarillo-Herrero,  Pablo and Ashoori,  Raymond},
  year = {2024},
  month = July,
  pages = {53–56}
}

@article{Wang2022,
  title = {Interfacial ferroelectricity in rhombohedral-stacked bilayer transition metal dichalcogenides},
  volume = {17},
  ISSN = {1748-3395},
  url = {http://dx.doi.org/10.1038/s41565-021-01059-z},
  DOI = {10.1038/s41565-021-01059-z},
  number = {4},
  journal = {Nature Nanotechnology},
  publisher = {Springer Science and Business Media LLC},
  author = {Wang,  Xirui and Yasuda,  Kenji and Zhang,  Yang and Liu,  Song and Watanabe,  Kenji and Taniguchi,  Takashi and Hone,  James and Fu,  Liang and Jarillo-Herrero,  Pablo},
  year = {2022},
  month = Jan,
  pages = {367–371}
}

@article{Jiang2024,
  title = {Vapor Deposition of Bilayer 3R MoS$_2$ with Room‐Temperature Ferroelectricity},
  volume = {36},
  ISSN = {1521-4095},
  url = {http://dx.doi.org/10.1002/adma.202400670},
  DOI = {10.1002/adma.202400670},
  number = {32},
  journal = {Advanced Materials},
  publisher = {Wiley},
  author = {Jiang,  Hanjun and Li,  Lei and Wu,  Yao and Duan,  Ruihuan and Yi,  Kongyang and Wu,  Lishu and Zhu,  Chao and Luo,  Lei and Xu,  Manzhang and Zheng,  Lu and Gan,  Xuetao and Zhao,  Wu and Wang,  Xuewen and Liu,  Zheng},
  year = {2024},
  pages = {2400670}
}

@article{Yang2023,
  title = {Ferroelectric transistors based on shear-transformation-mediated rhombohedral-stacked molybdenum disulfide},
  volume = {7},
  ISSN = {2520-1131},
  url = {http://dx.doi.org/10.1038/s41928-023-01073-0},
  DOI = {10.1038/s41928-023-01073-0},
  number = {1},
  journal = {Nature Electronics},
  publisher = {Springer Science and Business Media LLC},
  author = {Yang,  Tilo H. and Liang,  Bor-Wei and Hu,  Hsiang-Chi and Chen,  Fu-Xiang and Ho,  Sheng-Zhu and Chang,  Wen-Hao and Yang,  Liu and Lo,  Han-Chieh and Kuo,  Tzu-Hao and Chen,  Jyun-Hong and Lin,  Po-Yen and Simbulan,  Kristan Bryan and Luo,  Zhao-Feng and Chang,  Alice Chinghsuan and Kuo,  Yi-Hao and Ku,  Yu-Seng and Chen,  Yi-Cheng and Huang,  You-Jia and Chang,  Yu-Chen and Chiang,  Yu-Fan and Lu,  Ting-Hua and Lee,  Min-Hung and Li,  Kai-Shin and Wu,  Menghao and Chen,  Yi-Chun and Lin,  Chun-Liang and Lan,  Yann-Wen},
  year = {2023},
  month = Nov,
  pages = {29–38}
}

@article{Zheng2020,
  title = {Unconventional ferroelectricity in moiré heterostructures},
  volume = {588},
  ISSN = {1476-4687},
  url = {http://dx.doi.org/10.1038/s41586-020-2970-9},
  DOI = {10.1038/s41586-020-2970-9},
  number = {7836},
  journal = {Nature},
  publisher = {Springer Science and Business Media LLC},
  author = {Zheng,  Zhiren and Ma,  Qiong and Bi,  Zhen and de la Barrera,  Sergio and Liu,  Ming-Hao and Mao,  Nannan and Zhang,  Yang and Kiper,  Natasha and Watanabe,  Kenji and Taniguchi,  Takashi and Kong,  Jing and Tisdale,  William A. and Ashoori,  Ray and Gedik,  Nuh and Fu,  Liang and Xu,  Su-Yang and Jarillo-Herrero,  Pablo},
  year = {2020},
  month = Nov,
  pages = {71–76}
}

@article{Song2026,
   author = {Seunguk Song and Michael Altvater and Wonchan Lee and Hyeon Suk Shin and Nicholas Glavin and Deep Jariwala},
   doi = {10.1038/s44287-025-00251-w},
   issue = {3},
   journal = {Nature Reviews Electrical Engineering},
   month = {1},
   pages = {185-201},
   publisher = {Springer Science and Business Media LLC},
   title = {Indium selenides for next-generation low-power computing devices},
   volume = {3},
   year = {2026}
}

\newpage

\end{document}